# Usage-Based Vehicle Insurance: Driving Style Factors of Accident Probability and Severity

K. Korishchenko[1], I. Stankevich[2], N. Pilnik[3], D. Petrova[4]

The paper introduces an approach to telematics devices data application in automotive insurance. We conduct a comparative analysis of different types of devices that collect information on vehicle utilization and driving style of its driver, describe advantages and disadvantages of these devices and indicate the most efficient from the insurer's point of view.

The possible formats of telematics data are described and methods of their processing to a format convenient for modelling are proposed. We also introduce an approach to classify the accidents' strength. Using all the available information, we estimate accident probability models for different types of accidents and identify an optimal set of factors for each of the models. We assess the quality of resulting models using both in-sample and out-of-sample estimates.

1. **Introduction**

Internet of Things (IoT) is currently rising not only in consumer sphere, but also in application to different business processes. Combined with modern data analysis and data mining techniques it can significantly enhance the performance of various business automation and decision support systems, including, among other applications, the use of on-line individual-level data in insurance (Usage-Based Insurance). Our study is focused on one of these applications: the use of telematics data in vehicle insurance.

At the moment, the general approach to vehicle insurance is the use of commonly available factors, such as vehicle characteristics and general information about the driver (age, gender, driving experience, accident history). One of the ways to enrich this data is to add driving style indicators that cannot be measured directly, but that can be estimated using the data on how and when the driver uses his car.

The first approaches to application of telematics data to vehicle insurance are presented, e.g., in Husnjak et al., 2015, but utilize a smaller set of factors then proposed in our paper. A more thorough analysis is performed in Baecke, Bocca, 2017, with a considerably wide set of models (including not only econometrics models, but also machine learning approaches) and accuracy comparison of models with different sets of indicators. Bian et al., 2018 provides a more descriptive overview of Usage-Based systems in vehicle insurance.

Another very close and significantly better-researched problem is accident analysis. The current research in accident analysis and modeling is mostly centered on road- or city-level data, where aspects of road design and traffic organization are discussed or accident-level data, where accident participants' characteristics as well as some external factors, such as weather conditions, are considered. The first group is represented, for example, by Quddus, 2008, which uses time series econometrics models to predict the number of accidents in London and Great Britain; Andersen, 2007, Mohaymany et al., 2013, Bil et al., 2013 identify the spots with high accident probability on road maps. Erdogan et al., 2008 considers climatic, geographic and day-of-week factors. Olszewski et al. 2015 also discusses the influence of time period, area and road types and lightning conditions on the occurrence of accidents with pedestrians.

---

[1] Russian Presidential Academy of National Economy and Public Administration
[2] National Research University Higher School of Economics; The Lebedev Physical Institute of the RAS (LPI RAS)
[3] National Research University Higher School of Economics; Federal Research Center "Computer Science and Control" of the RAS,; The Lebedev Physical Institute of the RAS (LPI RAS)
[4] Meta Telematics

These studies give some insights into external factors influencing road accidents, such as weather conditions and road design, but they do not give any information on how driver-related factors influence accident probability. Studies based on accident-level data are more fruitful in this sense. Tesema et al., 2005 analyzes the influence of driver (age, gender, driving experience) and vehicle characteristics (age and type) as well as common external factors, such as weather and lightning conditions on accident severity. A similar approach is adopted in Kashani et al., 2011 and Kashani et al., 2012.

Studies discussing the influence of personal characteristics and driving behavior are scarce and are mostly based on survey data. There are some psychological studies on the connection between character traits and driving behavior. For example, French et al., 1993 analyzes the relationship between decision-making characteristics of a person and her driving behavior (both are based on specialized questionnaires) using a sample of 711 persons of different age and gender. Elliott et al., 2003 studies the influence of socio demographic characteristics and personal traits on the intention to violate the speed limits. Taubman-ben-Ari, 2004 uses specialized questionnaires to construct driving style indicators and studies the relationship between these indicators and personality traits. But, the data used in these studies is quite aggregated, based on subjective indicators and may correspond with actual driving behavior poorly.

More detailed data partly similar to what we use is considered in Park et al., 2017. Some indicators of driver's behavior, such as alcohol use, speeding and signal violation are used to model the severity of an accident. Kosuge et al., 2017 predicts the crash involvement of elderly drivers using driver characteristics based on survey data.

The structure of the rest of the paper is as follows. Chapter 2 provides an overview of Usage-Based Insurance systems and describes the model of telematics devices application in insurance industry. We describe the possible options for the target functions of the insurance company depending on the need for customer feedback. We also show that accounting for accidents can have a multi-level structure, which takes into account not only the fact of the accident, but also the degree of its severity. Chapter 3 describes the data collected by telematics devices and proposes an approach to calculation of a limited and interpretable set of indicators that can be used for scoring based on the original raw data. Chapter 4 presents some modelling results of car accident probability based on these indicators.

## 2. Usage-Based Insurance and Telematics

Historically classic Motor Own Damage (MOD) insurance business is based on actuarial calculations (customer's age, gender, driving experience, address and claims history, value, production year, brand and model, type and category, engine power of car, number of drivers, Stolen Vehicle Recovery and Stolen Vehicle Tracking (SVR/SVT) systems embedded etc.) to optimize the ratio of insurance premiums to insurance claims. Such approach allows to build 1-year insurance programs, because model of risk probability is calculated on year-base as minimum level of forecast.

There is a need from insurance business to build discrete actuarial models of different frequencies (1/3/6/12/36 month etc.). New technologies have already brought to the market the range of telematics devices, which gather driving statistics and behavioral data and detect and reconstruct crashes in real-time. Insurers use telematics-based pricing to reduce risks. Balance must be obtained between the acceptability of the pricing and the absolute effect of a parameter on losses. What telematics brings to Insurers is abbreviated as RAID: Real Time data (vs. historical); Actual Data (vs. statistical), Individual Data (vs. average of a risk class), Dynamic Data (vs. static). It would be more appropriate to know additionally how, when, where and how much the end user is driving the car. One of the key approaches here is Usage-Based Insurance, where insurance premiums are not fixed but rather determined by the mode of use.

Usage-Based Insurance (UBI) Programs can perform several functions simultaneously:

- detection and prevention of various fraud (TBYB - Try-Before-You-Buy - and pre-selection),
- decrease of combined ratio (the sum of an expense ratio and a loss ratio),
- reduction of the cost of claims management (portfolio segmentation),
- fuel consumption and car wear control (using so called CAN-bus OBD Devices - Controller Area Network On-Board Diagnostic Devices),
- improvement of actuarial models,
- increase of customer retention and loyalty,
- provision of individual feedback.

In long run individual Discounts on Insurance Premiums, no payment for the OBU (On-Board (telematics) Unit/Black Box), free Mobile App, the opportunity to easily use remote claims management, Ecall, Bcall, SVR etc. services for the End Users lead to UBI policies sales growth. UBI Scoring Solutions are used for pre-selection, anti-fraud and segmentation of the insurance portfolio.

In fact, UBI is a notion that includes all policies that make the premium dependent on the level of usage by drivers. UBI programs adapt insurance premiums to actual risk level. Telematics can help to build predictive models for forecasting of the future risk level of both the driver and insurance portfolio. Thanks to UBI, the actuary has more rating factors and that some of these have a higher predicting power than statistical factors. By correlating this driving behavior with actual claims, the insurer can, for each customer, establish the basic following formula much more accurately:

Insurance premium = Accident probability *Predicted Loss + Administrative costs + Margin

This outcome can have a significant impact on insurer's Profit and Loss account.

Based on PTOLEMUS research, 2016 UBI is becoming mainstream in auto insurance worldwide, especially in the US, Italy, UK etc. The US will become the leading UBI market in the world. In Europe, growth will be driven by Italy but the UK, Germany and France will see UBI subscriptions take off in the next 5 years. New major markets will emerge, including China and Russia. By 2020, nearly 100 million vehicles globally will be insured with telematics policies. This will grow to nearly 50% of the world's vehicles by 2030, generating more than €250 billion in premiums for insurers. In the US and Europe, most carmakers will have adopted UBI by 2020. The most successful model will use central data hub provided by Telematics Service Provider (TSP) connected to insurance companies. OBD dongles will become the leading UBI device, reaching all the continents. Aftermarket black boxes will continue to grow, specifically in high premium markets and for high value cars. Smartphone-based PHYD programs are set to become one of the key devices to collect driving data, with or without a Bluetooth beacon. However, they will not replace dongles or black boxes, which will still make up the majority of devices used in 2020.

Telematics data can be generated from a number of different sources:

- The car network, in particular the CAN-bus, part of which is accessed by a dongle through the OBD port,
- Sensors embedded in the vehicle, generally including motion and location sensors but also more commonly magnetometers (which can help determine the vehicle direction)
- Mobile devices, notably smartphones, which include all the sensors but are not physically linked to the vehicle.

While the car network and fixed devices can identify the car (thanks to VIN) and bring highly accurate data on braking, acceleration and car maneuvers, the telematics device must connect to the OBD interface to listen to the on-board diagnostics parameters ID (PID). Many are standard, defined by the SAE's J/1979 standard. However OEM's (Original Equipment Manufacturer) all define other non-standard PID. There is no standard telematics data, even Can-bus data is not standardized by OEM's yet.

Telematics devices, notably black boxes, are also often fitted with one or sometimes two accelerometers (generally with 3 axes). While fixed devices measurements are generally more reliable, they have now to compete with smartphone-based sensors that includes magnetometers (compasses) and microphones which can help detect the type of road, occupancy and the type of car driven.

For the UBI service provider, the choice of data is great but this data is only useful until the most predictive criteria is identified. The dataset selected will therefore change with experience and the understanding of what really identifies risky behavior and which behavior is most predictive of losses.

Opinions on what data is predictive are split and each insurer will have a different perspective on the weight of each criteria. Generally, the most mentioned are mileage, harsh braking and time of day. But predictability is not linear and the true values comes when sensor data is augmented by:

- Historical data: Motor Vehicle Record (MVR), claims history, CRM data,
- Static factors: time, seasonality, crime rate in the area, geography,
- Dynamic data feeds such as weather or traffic,
- Map attributes including road class, maximum speed, number of lanes, etc.
- Places (contextual location) such as crossings, schools, licensed establishments.

Insurers will also have different priorities. Focusing on accident frequency and claims ratio are completely different strategies. Italy has a high frequency but low death ratio. South Africa has very high death ratio but low frequency.

Smartphone-based solutions including "White Label" ones help to run digital marketing campaigns in order to generate pre-selected sales funnel for any concrete groups of customers. By installing the telematics SDK into Insurance Company smartphone app they immediately get access to telematics data and insights. Due to cost-effective nature of the insurer and ever increasing penetration worldwide, smartphones has the potential to reach consumers on a mass scale in every segment of the market. The range of embedded components means that a smartphone can be used as both a tracking device and a data hub. In addition, secure payments and account management plug-ins are increasingly available on all smartphones. TBYB carries several benefits, including a potentially faster time-to-market, as it does not require the insurer to integrate the user in their customer lifecycle. The telematics app contains the only necessary information that helps drivers to be a part of community and get insights to drive safely. The application provides a direct marketing channel to the individual, leading to improved customer acquisition. Drivers are able to use the driving feedback app as means to improve their own driving skills. Driving challenges are an effective tool to attract potential new customers and develop greater brand awareness. Rewards are much easier to scale to the complete portfolio of customers.

Nevertheless, the availability of telematics data does not mean that it will be used in insurance companies automatically. The aim of this paper is to propose an approach of estimation the riskiness of client using the telematics data. As we will see later, it is impossible to obtain a model of sufficient accuracy without the use of specialized data. In our case, these are data on the use of accelerations of various types by the driver.

3. **Methodology and data description**
3.1. **Types of telematics devices**

UBI OBU's collect telematics data using GNSS positioning (GPS/GLONASS, etc.) and communicate to Server via GSM/GPRS (for beacons via SMS) modem with embedded global M2M e-SIM (2G/3G). Server receiver is getting and parsing (MXP, XML, JSON) raw data by UDP, providing OBU diagnostics, manage and updating device firmware (OTA). Than filtered and aggregated telematics data is being used in web/mobile products for business matters in UBI/FMS/SVR/SVT, etc.

E.g. Meta System UBI OBU's have Meta System1 proprietary MXPII protocol of communication and two internal 3-axis accelerometers with gyro integrated, one of which is used on driver behavior (2 g accelerations) and crash detection when the ignition is off, the other one – for crash detection and 3D road accident reconstruction (up to 24 g accelerations). The Bluetooth 4.0 open source protocol is used for all Meta product line to pair to Meta wide range of wireless accessories (ID Tag, E-Call, B-Call buttons, Engine Immobilizer, Car Alarm or OBD CAN-Bus dongle). OBD Dongles read vehicle CAN-Bus Data, e.g. Fuel consumption, Fuel Tank level, DTC, Malfunction Indicator Lamp, etc.

The telematics-enabled motor insurer has access to:

- Vehicle data, such as mileage, driving times, driver behavior's, phantom drivers, thefts, accidents, types of roads;
- Driver data, if there is only single driver insured; Experienced insurers will also be able to identify a new driver based on his/her driving style. Our research has confirmed that we can identify driving style of End User in a 3 months' period.
- Environment data, notably the type of roads (motorways, urban roads, etc.), traffic jams, weather, urban/rural environment, dangerous spots.

These data sets can represent around 2-45 Mb per month per customer, depending on the list of indicators collected, the frequency and the number and length of annual trips. Properly collecting the data that matters is crucial to obtain the accurate scoring.

### 3.2. Original data format

Originally, the data is received as a set of data packages from the telematics device that can contain different indicators. Some of them provide purely technical information about the device functioning, such as voltage, different test results etc. More informative packages contain information about the ignition being turned on and off, vehicle coordinates, speed and accelerations.

Packages are sent not on regular time intervals, but when some predefined events take place. This approach seems natural for ignition-related events. Acceleration-related events take place when the acceleration of one of the axis is higher then some threshold value in order to save only the most informative data. Speed-related events take place when the vehicle drives with speed above some threshold. This approach allows for more compact yet information-preserving data storage but makes the data processing more complicated because of the need to account for differing data frequency. The number of packages obtained from a vehicle with the current settings is around several thousand per day.

Time aggregated data is more convenient for the purposes of driving behavior description. In this case all the packages data is turned to data of certain frequency (hourly, daily, weekly, monthly and so on) and indicators used are turned to mean of cumulative values during the corresponding time period. Our research shows that aggregation to the hourly data is the most convenient, hence, instead of separate events we use cumulative mileage over an hour, mean speed and mean acceleration of different strength frequencies.

It allows us to move from several thousand observations per car per day to 24 observations per day (see figure 1 for sample data). As the vehicle is typically not used 24 hours per day some observations are zero (figure 2). The situation also changes because of morning and evening rush hour traffic congestions and the effect of weekend and holidays.

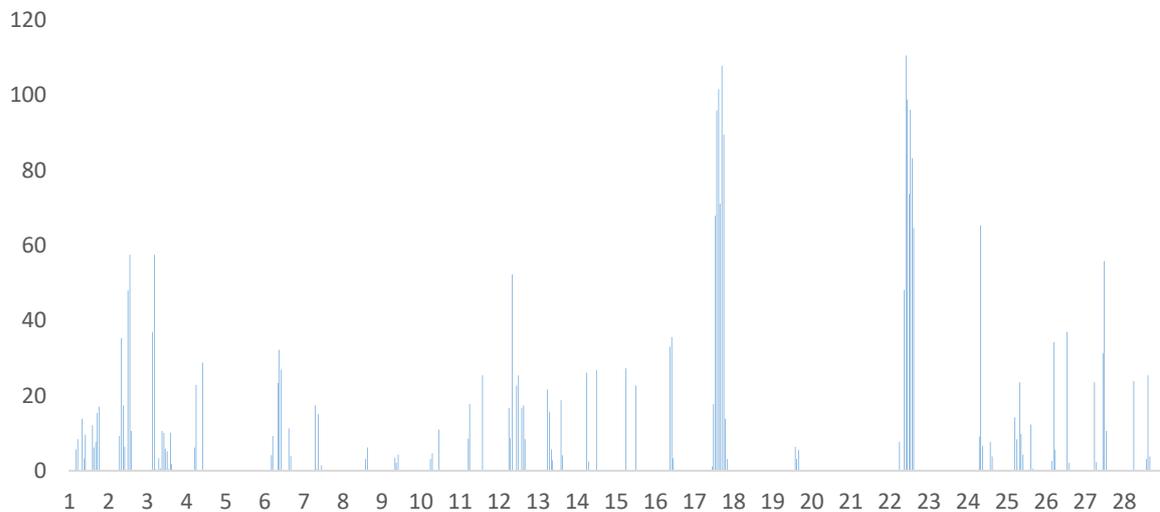

Figure 1. Hourly car mileage during February 2019, km. Day number on horizontal axis

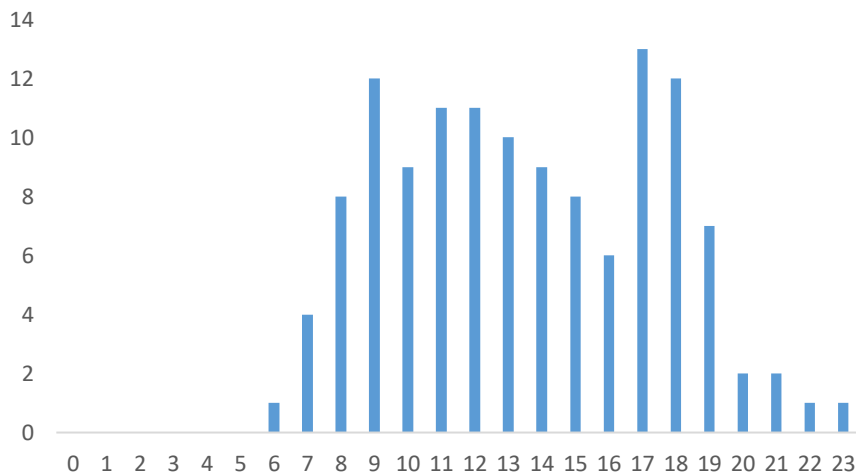

Figure 2. Number of days in February 2019 with non-zero mileage for a particular hour. The hour number on horizontal axis.

### 3.3. Data aggregation and feature engineering

On the one hand, the obtained dataset describes the driver's behavior during the day in a detailed and very informative way. On the other hand, this information is not structured enough to be used in scoring algorithms.

To solve this contradiction, we need to determine the format of data aggregation. In our case we need to make two transformations. Firstly, we move from hourly intervals to longer time intervals. The research has shown that the two types of most informative intervals are weekly intervals and all period of observation intervals. The first type of data describes the current behavior of the driver and provides an immediate picture of driving style. The second allows to analyze the driving style in a long run perspective.

Secondly, we introduce a wider range of indicators that characterize the mileage, speed and accelerations to preserve the information on driving style differences between business days and weekends and different times of day. We add the following list of mileage indicators:

- mileage – total mileage during the observation period (in kilometers)
- trips_day – average number of trips per day
- below_10_pr – share of trips shorter than 10 km
- below_30_pr – share of trips shorter than 30 km
- over_200 – share of trips longer than 200 km
- over_400 – share of trips longer than 400 km
- d_total_m – average mileage
- avg_trip_mil – average trip mileage
- avg_trip_dur – average trip duration
- d_business_m – average mileage on business days
- d_day_m – average daytime mileage (7 am – 7 pm)
- d_evening_jam_m – average evening rush hour mileage (18 am – 20 am)
- d_morning_jam_m – average morning rush hour mileage (8 am – 10 am)
- d_holi_m – average holiday mileage
- d_night_m – average nighttime mileage (0 am – 6 am)
- day_m_pr – ratio of daytime mileage to total mileage
- ej_m_pr – ratio of evening rush hour mileage to total mileage

Figures 3-4 show the dynamics of weekly mileage and average morning rush hour weekly mileage averaged for all cars in the dataset. We can clearly see the seasonal wave with summer period and New Year week (official holiday week in Russia)

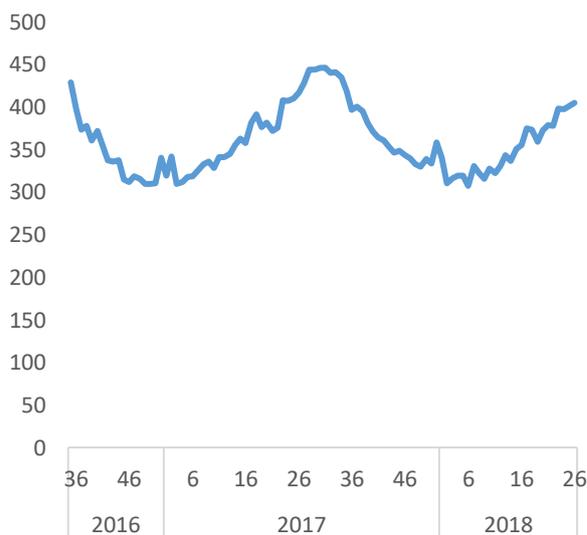

Figure 3. Average weekly mileage, km.

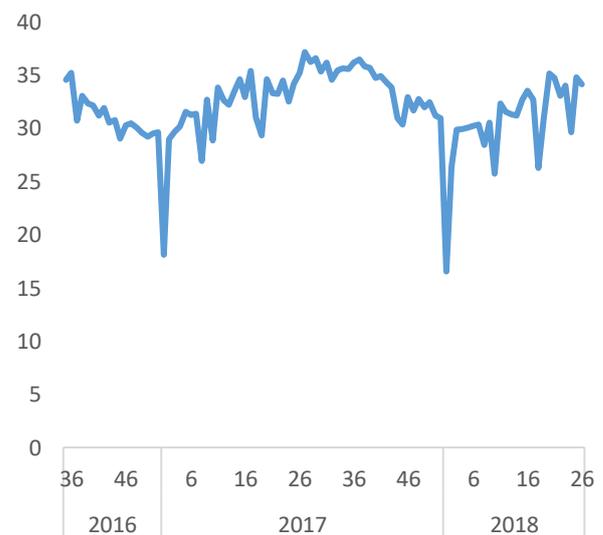

Figure 4. Average weekly morning rush hour mileage, km.

The following speed indicators are added to the dataset:

- avg_sp – average speed,
- max_sp – maximum speed (over the whole observation period)
- max_ej_sp – maximum speed during evening rush hour
- max_mj_sp – maximum speed during morning rush hour
- max_n_sp – maximum night speed
- m_pr_below_20 – share of mileage with speed lower than 20 kph
- m_pr_below_60 – share of mileage with speed lower than 60 kph
- m_pr_over_100 – share of mileage with speed over 100 kph
- m_pr_over_130 – share of mileage with speed over 130 kph.

Figures 5-6 show the dynamics of average speed and average evening rush hour speed averages for all the cars in the dataset. The seasonal wave is also clear here

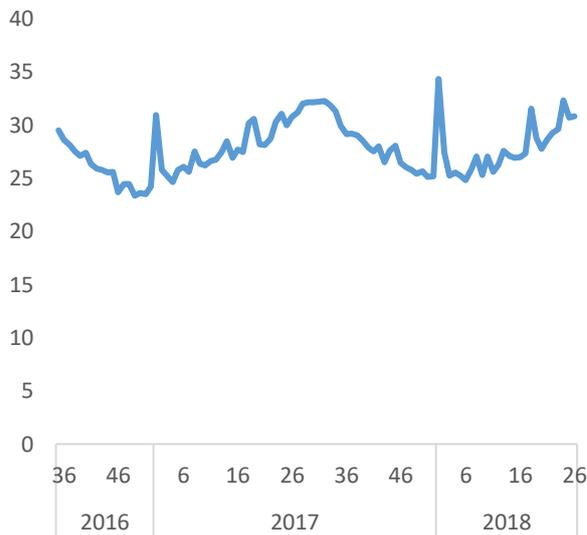 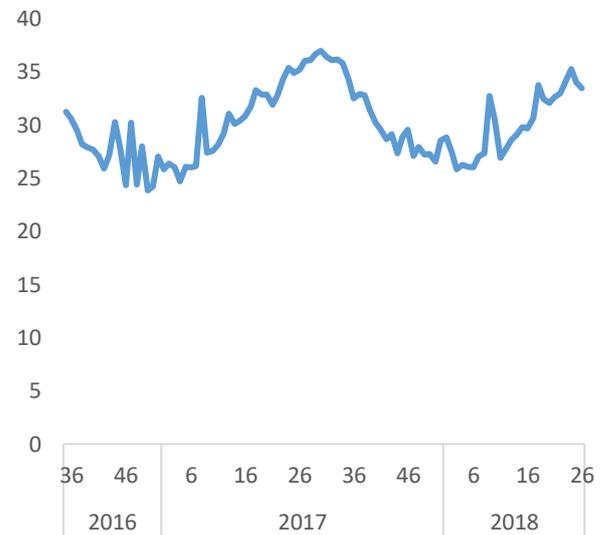

Figure 5. Average weekly speed, kmh    Figure 6. Average evening rush hour speed, kmh

The accelerations data is more complicated: simple calculation of the number of accelerations above certain threshold does not contain any information about direction and strength of these accelerations. To preserve these information in the final dataset we consider the following indicators separately:

- Accelerations (the positive component of acceleration along the longitudinal axis)
- Decelerations (the negative component of acceleration along the longitudinal axis)
- Side accelerations (absolute value of acceleration along the transverse axis)

For every indicator we introduce three levels of strength and calculate the frequency of corresponding accelerations, decelerations and side accelerations per 100 kilometers to eliminate the effect of mileage. The resulting set of indicators is as follows:

- a1 – frequency per 100 km of level 1 acceleration (0.3 – 0.4 G ) events
- a2 – frequency per 100 km of level 2 acceleration (0.4 – 0.5 G) events
- a3 – frequency per 100 km of level 3 acceleration (0.5+ G) events
- d1 – frequency per 100 km of level 1 deceleration (0.2 – 0.3 G) events
- d2 – frequency per 100 km of level 2 deceleration (0.3 – 0.4 G) events
- d3 – frequency per 100 km of level 3 deceleration (0.5+ G) events

- s1 – frequency per 100 km of level 1 side acceleration (0.3 – 0.4 G) events
- s2 – frequency per 100 km of level 2 side acceleration (0.4 – 0.6 G) events
- s3 – frequency per 100 km of level 3 side acceleration (0.6+ G) events

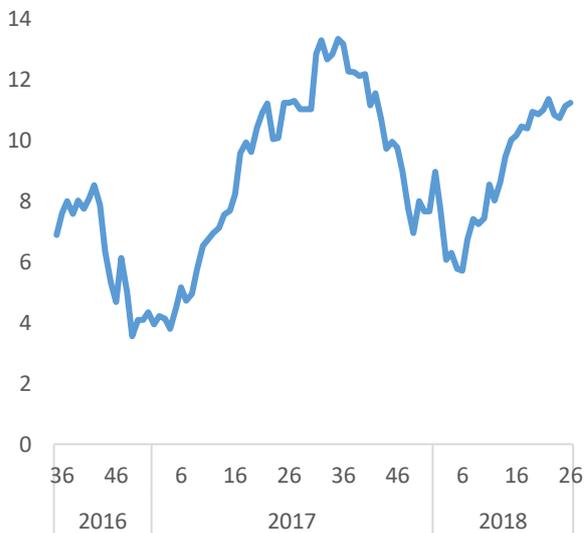

Figure 7. Average frequency of level 1 accelerations, events per 100 km.

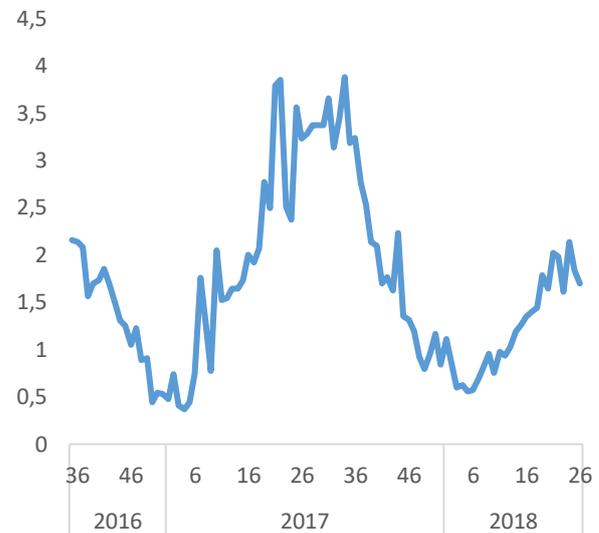

Figure 8. Average frequency of level 2 side accelerations, events per 100 km.

The resulting dataset contains the following groups of indicators:

- Mileage indicators

Mileage indicators, even though specific for each driver, mostly characterize the utilization of the vehicle, not the driving style itself. In medium run perspective (several months to several years) a driver usually cannot influence the distance between the main points she is located during the day (mostly home and work). Hence, there is always some minimum daily mileage. The same situation is with timing of her mileage that is determined by working hours. Both mileage and its structure do influence the accident probability, but one should understand the mechanism behind this influence.

- Speed indicators

Speed indicators, especially average speed indicators, also characterize utilization of the vehicle and external conditions of her trips rather than the driving style. Mean speed and mean speed at different times of day and weekdays is determined mostly not by the driving style of the driver but by the situation on roads and traffic congestions. At certain moments of time the driver can choose less congested roads, but, generally, even in this case he is restricted by the characteristics of place he lives in. Road congestions influence the accident probability, but it is not determined by the driver's behavior.

- Acceleration indicators

In contrast to the previous groups of indicators, acceleration indicators mostly characterize the driving style of a particular driver. Intensities of accelerations, decelerations and side accelerations (that characterize cornering behavior and turning style) can be used as indicators of driving style. These information allows both to obtain a description of medium run and long run driving style and to track the changes in behavior inside a week. The availability of acceleration indicators is the key feature of our data that is possible to obtain using telematics devices, that distinguishes our data from what is usually used. As shown below, this unique information can significantly improve the quality of resulting models.

### 3.4. Accident rate in the data

To perform a more thorough analysis of factors influencing the accident probability it was decided to separate the accidents into 3 groups accordingly to the ratio of losses over the observation period to the insurance sum. Analysis of the fig. 2 allows us to form 3 groups (vertical lines on the plot are borders of the groups):

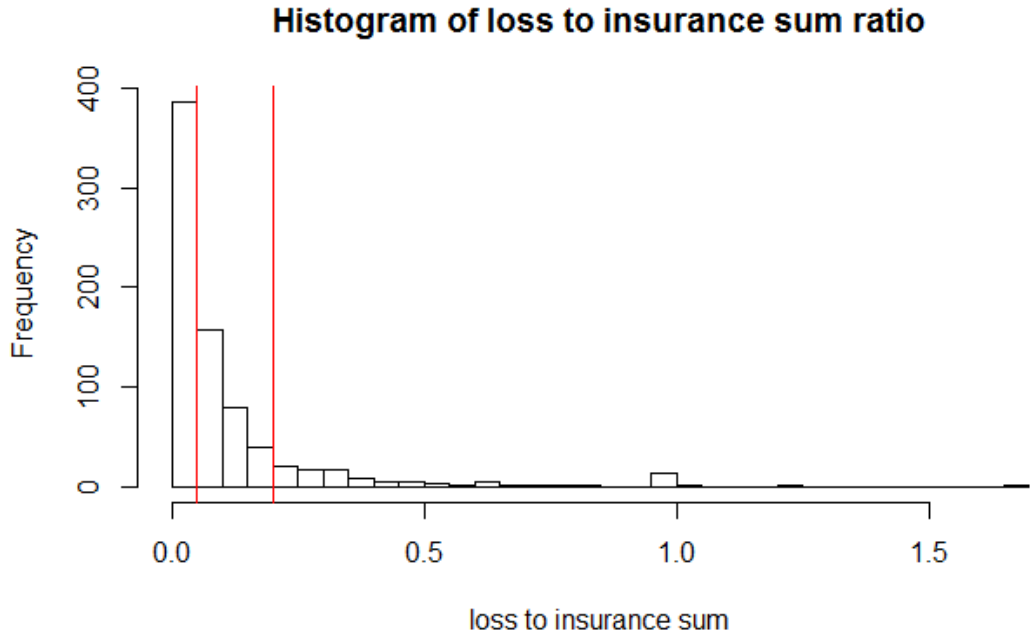

*Figure 9: histogram of loss to insurance sum ratio for accident culprits*

It was decided to treat accidents with loss to insurance ratio smaller than 5% as small, from 5% to 20% as medium-sized, greater than 20% as strong. In some cases the same vehicle has more than one accident, hence, there can be a bias of some sort (for example, we cannot distinguish between two small accidents and one medium), but the number of observations with more than one accident is small and this factor will not influence the results significantly. This approach gives us 386 observations with weak accidents, 278 with medium and 102 observations with strong accidents. We also have 30 observations with insurance company mark that a vehicle was in an accident but with zero losses, we treat them as observations without accident (probably, there was an accident, but a minor one and no repairs was needed, so, we can neglect it).

### 3.5. Some descriptive statistics

Some descriptive statistics are presented in table 1

| Independent variables | Drivers with accidents | | Drivers without accidents | |
|---|---|---|---|---|
| | Mean | std | Mean | std |
| mileage | 14561.17 | 11507.88 | 9238.27 | 9324.56 |
| trips_day | 7.77 | 71.68 | 4.98 | 14.95 |
| sp1 | 0.0016 | 0.0018 | 0.0013 | 0.0017 |
| sp2 | 0.0002 | 0.0005 | 0.0002 | 0.0004 |
| sp3 | 0.0000 | 0.0001 | 0.0000 | 0.0001 |
| a1 | 20.01 | 17.74 | 16.63 | 19.46 |
| a2 | 2.16 | 3.98 | 2.40 | 12.13 |
| a3 | 1.69 | 39.96 | 1.45 | 37.82 |
| d1 | 4.73 | 8.32 | 5.99 | 21.99 |
| d2 | 0.70 | 2.28 | 1.37 | 25.14 |
| d3 | 0.55 | 14.70 | 3.38 | 80.15 |
| s1 | 5.76 | 10.85 | 5.72 | 14.06 |
| s2 | 0.65 | 1.49 | 1.08 | 17.18 |
| s3 | 1.13 | 20.46 | 1.59 | 35.18 |
| below_10_pr | 68.63 | 16.48 | 70.69 | 17.53 |
| below_30_pr | 90.56 | 9.84 | 91.87 | 9.46 |
| over_200 | 4.74 | 8.58 | 2.99 | 6.71 |
| over_400 | 0.58 | 1.58 | 0.40 | 1.28 |
| avg_sp | 24.88 | 8.27 | 24.68 | 8.72 |
| avg_trip_mil | 13.12 | 7.27 | 12.15 | 7.85 |
| avg_trip_dur | 1926.68 | 1023.17 | 1809.09 | 1395.79 |
| d_business_m | 63.36 | 183.82 | 47.96 | 43.41 |
| d_day_m | 51.59 | 211.29 | 37.99 | 31.35 |
| d_evening_jam_m | 7.63 | 29.19 | 5.79 | 7.98 |
| d_morning_jam_m | 11.48 | 41.04 | 8.35 | 10.68 |
| d_holi_m | 59.73 | 126.19 | 46.18 | 40.56 |
| d_night_m | 13.71 | 34.95 | 9.35 | 14.17 |
| d_total_m | 65.76 | 243.76 | 47.80 | 40.78 |
| day_m_pr | 78.18 | 12.99 | 80.89 | 13.59 |
| ej_m_pr | 8.36 | 7.47 | 8.57 | 8.37 |
| max_ej_sp | 122.42 | 42.77 | 107.50 | 44.19 |
| max_mj_sp | 134.22 | 36.23 | 119.05 | 39.37 |
| max_n_sp | 140.26 | 36.67 | 122.95 | 42.30 |
| max_sp | 161.57 | 45.70 | 146.13 | 36.99 |
| m_pr_below_20 | 13.53 | 6.34 | 14.03 | 8.09 |
| m_pr_below_60 | 49.00 | 16.44 | 50.93 | 18.47 |
| m_pr_over_100 | 15.13 | 12.72 | 13.65 | 12.86 |
| m_pr_over_130 | 2.70 | 5.07 | 2.01 | 4.23 |
| loss_size | 171006 | 533781 | - | - |
| ins_sum | 1728042 | 1276152 | | |

*Table 1: descriptive statistics*

It can be noted, that the means of some variables differ between the two groups, especially mileage and average number of trips per day, higher mileages and numbers of trips per day associated with drivers, involved in accidents. The same is true for all other indicators of mileage. Accident-free drivers are also characterized by lower maximum speeds and shares of mileage with high (over 100 and over 130 kph) speed. Typical loss size is about 10% of typical insurance sum.

Correlation matrix of the variables used in model selection is presented on fig. 10. The bigger the circle the stronger is the correlation. Blue circles stand for positive correlation, red ones – for negative.

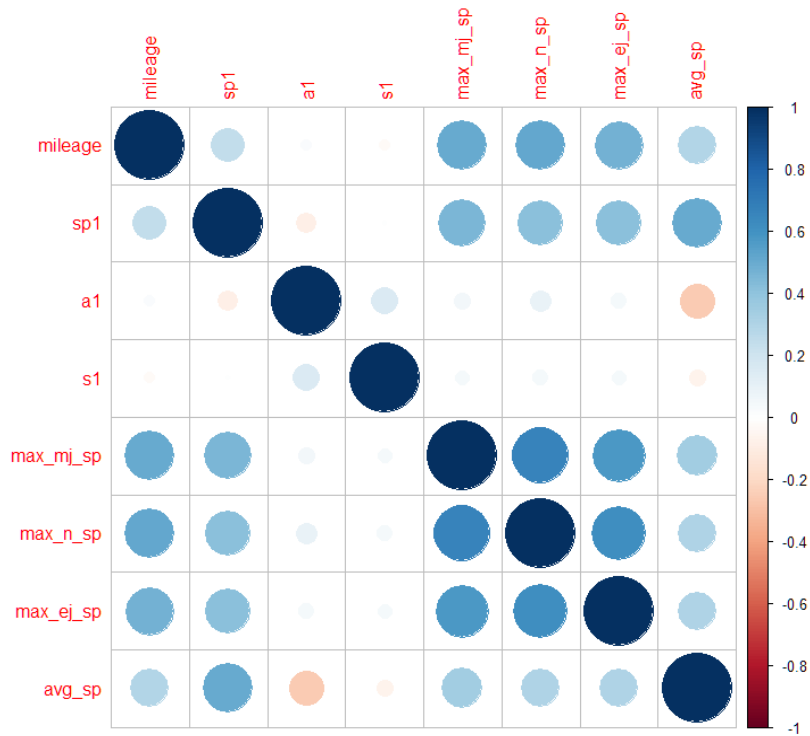

*Figure 10: correlation matrix for main variables*

We see that high maximum speeds are associated with each other and with high mileage. High frequency of 0-20 kph speed limit violations is, not surprisingly, also associated with maximum speeds as well as high average speed. Accelerations and side accelerations are generally independent of other variables, with only exception of negative correlation between 0.3-0.4G accelerations and average speed.

Correlation plot of all independent variables is presented on figure 11. The general tendencies noted for main variables, are also present for wider groups of indicators.

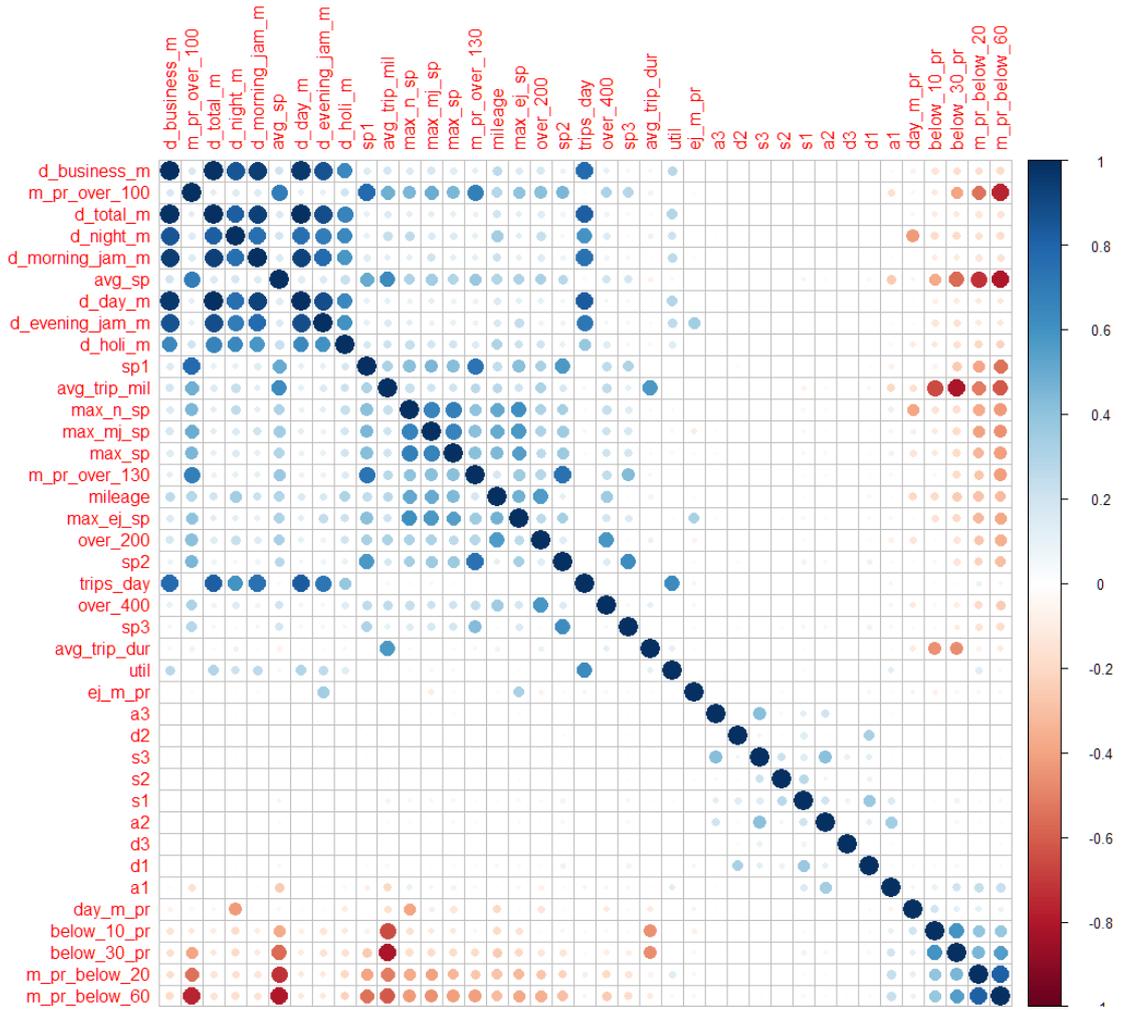

*Figure 11: Correlation matrix for all variables*

### 3.6. Modelling approach and accuracy evaluation

We use logistic regression as the main accident probability model. Here we follow the argumentation of Baecke, Bocca, 2017 that points out the importance of interpretability and necessity of "white box" models as a standard regulatory requirement.

Event probability in logistic regression is calculated as

$$P(Y_i \mid X_i) = \frac{1}{1 + exp^{\beta X_i}}$$

Where $X_i$ - vector of factors for i-th observation, including intercept; $\beta$ - vector of model parameters, estimated usually with maximum likelihood.

To measure the quality of resulting models we use in- and out-of-sample ROC AUC (area under the receiver operator characteristic curve) that can be interpreted as the probability that a randomly chosen positive instance (driver that has an accident) has probability higher then randomly chosen negative instance (driver without accidents). We also calculate more common in econometrics literature McFadden R-squared measure that is $R^2_{McFadden} = 1 - \frac{logL(model)}{logL(constant)}$, where $logL(model)$ and

$logL(constant)$ are logarithms of likelihood functions for the model under consideration and a model that gives a constant forecast.

## 4. Results and Discussion
### 4.1. A general approach

Before we proceed to accident probability models estimation and analysis it also important to describe our approach to data types when solving scoring problems. We have already mentioned the two approaches to data processing that differ in time interval that is used for data averaging and aggregation. Each of these approaches solves its own problem.

Lifetime-data based scoring (or long-term data based scoring) is used mostly to assess and filter the clients of an insuring company that uses the scoring system. There is no way to analyze short-term behavior of particular driver under this approach, but it is possible to make the general assessment of her driving style and skills.

Short-term scoring that uses weekly or monthly data (the work with daily data is much more complicated as it requires to deal with intra-week heterogeneity) can be used in purposes of monitoring and correction of current behavior of the driver. Under this approach it is possible to make a monitoring and individual recommendation system. In this paper we focus on the first approach.

In our sample we have 4284 observations without accidents, 597 observations with 1 accident and only 199 observations with more than 1 accident. Due to the small number of observations with more than 1 accident it was decided to restrict the model scope to simple logic with the fact of accident (whether the vehicle was in any accidents) as a dependent variable. We remind, that all the accidents discussed here are the accidents where the driver of the vehicle under consideration was the culprit of the accident.

4 models were estimated (for all accidents, and three groups separately), only statistically significant variables were left in models. As we are analyzing the binary outcome variable, we focus on logistic regressions. The table below contains the lists of significant factors for every accident type and the direction of their influence.

|            | All Accidents | Weak Accidents | Medium Accidents | Strong Accidents |
|------------|---------------|----------------|------------------|------------------|
| MILEAGE    | +             | +              | +                |                  |
| A1         | +             | +              | +                | +                |
| A2         | -             |                |                  | -                |
| AVG_SP     | -             | -              |                  |                  |
| MAX_N_SP   | +             |                | +                | +                |
| MAX_MJ_SP  | +             | +              |                  |                  |
| S1         |               | -              |                  | +                |
| D_NIGHT_M  |               |                | +                |                  |
| MAX_EJ_SP  |               |                |                  | +                |

*Table 2: The influence of individual factors on the probability of an accident*

As was anticipated, one of the main factors influencing the probability of accident is mileage. The higher the mileage, the higher is the probability of accidents – this rule holds for all models except the model for strong accidents – here only driving-style and car utilization factors matter. Frequency of 0.2-0.3G accelerations also increases the accident probability – again, everywhere except the model for strong accidents. 0.2-0.3G is the level achieved when the car starts to move – whether at traffic lights or in congestions. Hence, this variable can be viewed as an indicator of frequent use of a car in the city and in congested roads, where the probability of small accidents is higher (at least in Russia). The same is true for day_m_pr (ratio of daytime mileage to total mileage) – it reduces the probability of strong accidents (strong accidents more often happen at nighttime, this idea is confirmed by positive influence of average nighttime mileage on medium-sized accidents). The last car utilization indicator is average speed – the higher the average speed the less time the vehicle spends in traffic congestions, which reduces the probability of accidents in general and weak accidents.

Coming to driving style indicators we start with maximum speed during morning rush hour, night, evening rush hour increases the probability of weak, medium and strong accidents respectively (for first two cases the influence is also present in general model). The mechanism is quite clear – the higher the maximum speed, the less careful and more prone to accidents is the driver. The difference of influences of different types of maximum speeds is less clear and may be due to imperfections of data. At last, 0.2-0.3 G side accelerations (usually – sharp changes of lane) reduce the probability of weak accidents and increase the probability of strong ones. Weaving is typical for careless drivers, and the model estimates confirm that they are more prone to get into more serious accidents.

Now we move to detailed analysis of models separately and describe a portrait of typical driver that gets into an accident of certain type.

### 4.2. Detailed modelling results

**Results**

|  | *Dependent variable:* | | | |
|---|---|---|---|---|
|  | DTP_f | | | |
|  | All accidents | Week accidents | Medium accidents | Strong accidents |
| mileage | 0.0000*** | 0.0000*** | 0.0000*** | |
|  | (0.000) | (0.000) | (0.000) | |
| max_ej_sp |  |  |  | 0.007** |
|  |  |  |  | (0.003) |
| a1 | 0.010*** | 0.007** | 0.006** | 0.022*** |
|  | (0.003) | (0.003) | (0.003) | (0.006) |
| a2 | -0.029** |  |  | -0.119*** |
|  | (0.013) |  |  | (0.042) |
| max_mj_sp | 0.004*** | 0.009*** |  |  |
|  | (0.001) | (0.002) |  |  |
| s1 |  | -0.047*** |  | 0.017*** |
|  |  | (0.013) |  | (0.005) |
| avg_sp | -0.020*** | -0.021*** |  |  |
|  | (0.006) | (0.007) |  |  |
| max_n_sp | 0.005*** |  | 0.004** | 0.005* |
|  | (0.001) |  | (0.002) | (0.003) |
| d_night_m |  |  | 0.004* |  |
|  |  | (0.002) |  |  |
| Constant | -2.880*** | -3.352*** | -3.863*** | -5.641*** |
|  | (0.196) | (0.249) | (0.229) | (0.388) |
| Observations | 5,050 | 5,050 | 5,050 | 5,050 |
| Log Likelihood | -2,080 | -1,303 | -1,038 | -480 |
| Akaike Inf. Crit. | 4,174.6 | 2,618.3 | 2,087.4 | 972.7 |

Note: *p **p ***p<0.01

Table 3: The results of the estimation of the probability of accident models

A typical driver with high accident probability:

- Has high mileage,
- Makes a lot of of type 1 accelerations,
- Drives with low average speed,
- Drives fast at nights and mornings.

A typical driver with high week accident probability:

- Has high mileage,
- Drives with low average speed,
- Drives fast in the mornings,
- Makes a lot of type 1 accelerations but few type 1 side accelerations.

A typical driver with high medium accident probability:

- Has high mileage and high night mileage,
- Makes a lot of of type 1 accelerations,
- Drives fast at night.

A typical driver with high strong accident probability:

- Drives fast at evening,
- Drives a lot at night,
- Makes a lot of of type 1 accelerations and type 1 side accelerations,
- Makes few type 2 accelerations.

### 4.3. Accident probability models quality estimation

To estimate the forecasting power of the model we use one of the standard indicators: the area under the ROC curve (ROC AUC). The table 4 contains both the in-sample and out-of-sample results for the 4 models. For out-of-sample calculation, we used randomly chosen 10% of observations: the model was estimated on 90% of the sample, the forecast was built for the other 10% of observations.

|  | All Accidents | Weak Accidents | Medium Accidents | Strong Accidents |
|---|---|---|---|---|
| In sample | 0.676 | 0.673 | 0.681 | 0.696 |
| Out of sample | 0.659 | 0.650 | 0.693 | 0.691 |

Table 4: Comparison of the accuracy of the model inside and outside the sample

We can see that in-sample and out-of-sample estimates are very close. It may indicate that the models do not overfit the data and have a reasonable forecasting power.

Another important issue is the increase of models quality due to acceleration information introduction to the models. To measure it, we estimated the same models without acceleration indicators (with mileage and speed indicators as main variables). The table 5 compares McFadden R-squared for models with and without accelerations

|  | All Accidents | Weak Accidents | Medium Accidents | Strong Accidents |
|---|---|---|---|---|
| Models with accelerations | 0.0523 | 0.0441 | 0.0349 | 0.0373 |
| Models without accelerations | 0.0492 | 0.0367 | 0.0332 | 0.0195 |

*Table 5: Comparing model accuracy with and without acceleration*

We can see that introduction of acceleration indicators increases the quality of models, especially for week and strong accidents.

## 5. Conclusion

This paper contributes both to automotive insurance literature by offering a methodology of telematics data application from data collection and preprocessing to accident probability model estimation, and to accident research literature by providing some insights into driving style factors of accident probability and severity.

We present an application methodology for telematics devices data to vehicle insurance. The paper both discusses the general approach to data collection and preprocessing and identifies the most significant factors that influence the accident probability using real-world data. Both vehicle utilization and driving style indicators influence the accident probability, but the set of significant variables depends on the strength of the accident examined.

The results obtained in the paper can be valuable both for insurance companies as a guideline for telematics data application in automotive insurance and for authorities to prioritize the policy development.